\documentclass{bioinfo}
\copyrightyear{2014}
\pubyear{2014}

\usepackage{amsmath}
\usepackage[ruled,vlined]{algorithm2e}
\usepackage{natbib}

\begin{document}
\firstpage{1}

\title[Constructing BWT]{Fast construction of FM-index for long sequence reads}

\author[Li]{Heng Li}

\address{Broad Institute, 7 Cambridge Center, Cambridge, MA 02142, USA}

\history{Received on XXXXX; revised on XXXXX; accepted on XXXXX}
\editor{Associate Editor: XXXXXXX}
\maketitle

\begin{abstract}
\section{Summary:} We present a new method to incrementally construct the
FM-index for both short and long sequence reads, up to the size of a genome.
It is the first algorithm that can build the index while implicitly sorting the
sequences in the reverse (complement) lexicographical order without a separate
sorting step. The implementation is among the fastest for indexing short reads
and the only one that practically works for reads of averaged kilobases in
length.

\section{Availability and implementation:} https://github.com/lh3/ropebwt2

\section{Contact:} hengli@broadinstitute.org
\end{abstract}

\section{Introduction}

FM-index plays an important role in sequence alignment, {\it de novo}
assembly~\citep{Simpson:2012aa} and compression~\citep{Cox:2012ly}. Fast
and lightweight construction of FM-index for a large data set is the key to
these applications. In this context, a few
algorithms~\citep{DBLP:journals/tcs/BauerCR13,DBLP:journals/corr/LiuLL14} have
been developed for DNA sequences that substantially outperform earlier
algorithms.  However, they are only efficient for short reads and require
special hardware, a fast disk or a high-end GPU. An efficient and practical
algorithm for long sequence reads is still lacking. This work aims to fill this
gap.

\section{Methods}
\begin{methods}
Let $\Sigma=\{{\tt A},{\tt C},{\tt G},{\tt T},{\tt N}\}$ be the \emph{alphabet}
of DNA with a lexicographical order
${\tt A\char60C\char60G\char60T\char60N}$. Each element in $\Sigma$ is
called a \emph{symbol} and a sequence of symbols called a \emph{string} over
$\Sigma$. Given a string $P$, $|P|$ is its length and $P[i]$ the symbol at
position $i$. A sentinel $\$$ is smaller
than all the other symbols. For simplicity, we let $P[-1]=\$$ for any string
$P$. We also introduce $\widetilde{P}$ as the reverse of $P$ and
$\overline{P}$ as the reverse complement of $P$.

Given a list of strings over $\Sigma$, $(P_i)_{0\le i<m}$, let
$T=P_0\$_0\ldots P_{m-1}\$_{m-1}$ with
${\tt\char36}_0{\tt\char60}\cdots{\char60\tt\char36}_{m-1}{\tt\char60A\char60C\char60G\char60T\char60N}$.
The \emph{suffix array} of $T$ is an integer array $S$ such that $S(i)$,
\mbox{$0\le i<|T|$}, is the starting position of the $i$-th smallest suffix in
the collection $T$. The \emph{Burrows-Wheeler Transform}, or \emph{BWT}, of $T$
can be computed as \mbox{$B[i]=T[S(i)-1]$}. For the description of the
algorithm, we segment $B$ into \mbox{$B=B_{\tt\$}B_{\tt A}B_{\tt C}B_{\tt
G}B_{\tt T}B_{\tt N}$}, where \mbox{$B_a[i]=B[i+C(a)]$} with
\mbox{$C(a)=|\{j:T[j]<a\}|$} being the array of accumulative counts. By the
definition of suffix array and BWT, $B_a$ consists of all the symbols with
their next symbol in $T$ being $a$.

The above defines BWT for an order list of strings. We next seek to define BWT
for an unordered set of strings $\mathcal{C}$ by imposing an arbitrary sorting
order on $\mathcal{C}$.  We say list $(P_i)_i$ is in the \emph{reverse
lexicographical order} or \emph{RLO}, if $\widetilde{P}_i\le\widetilde{P}_j$
for any $i<j$; say it is in the \emph{reverse-complement lexicographical order}
or \emph{RCLO}, if $\overline{P}_i\le\overline{P}_j$ for any $i<j$.
The \emph{RLO-BWT} of $\mathcal{C}$, denoted by $B^{\rm RLO}(\mathcal{C})$,
is constructed by sorting strings in $\mathcal{C}$ in RLO and then applying
the procedure in the previous paragraph on the sorted list. \emph{RCLO-BWT}
$B^{\rm RCLO}(\mathcal{C})$ can be constructed in a similar way. In
$B^{\rm RCLO}(\{P_i\}_i\cup\{\overline{P}_j\}_j)$, the $k$-th smallest
sequence is the reverse complement of the $k$-th sequence in the FM-index. This
property removes the necessity of keeping an extra array to link the rank and
the position of a sequence in the FM-index, and thus helps to reduce the memory
of some FM-index based algorithms~\citep{Simpson:2012aa}. For short reads, RLO/RCLO-BWT
is also more compressible~\citep{Cox:2012ly}.

As a preparation, we further define two string operations: ${\rm
rank}(c,k;B)$ and ${\rm insert}(c,k;B)$, where ${\rm
rank}(c,k;B)=|\{i<k:B[i]=c\}|$ gives the number of symbols $c$ before the
position $k$ in $B$, and ${\rm insert}(c,k;B)$ inserts symbol $c$ after $k$
symbols in $B$ with all the symbols after position $k$ shifted to make room for
$c$. We implemented the two operations by representing the string $B$ with a
B+-tree, where a leaf keeps a run-length encoded string and an internal node
keeps the count of each symbol in the leaves descended from the node.

Algorithm 1 appends a string to an existing index by inserting each of its
symbol from the end of $P$. It was first described
by~\citet{DBLP:conf/cpm/ChanHL04}. Algorithm 2 constructs RLO/RCLO-BWT
in a similar manner to Algorithm 1. The difference lies in that it inserts $P[i]$ to
$[l,u)$, the suffix array interval of $P$'s suffix starting at $i+1$. This process implicitly
applies a radix sort from the end of $P$, sorting it into the existing strings
in the BWT in RLO/RCLO.  Note that if we change line 1 to ``\mbox{$l\gets
u\gets|\{i:B[i]=\$\}|$}'', Algorithm 2 will be turned into Algorithm 1. Recall
that the BCR algorithm~\citep{DBLP:journals/tcs/BauerCR13} is, to some extent, the
multi-string version of Algorithm 1. Following similar reasoning, we can extend
Algorithm 2 so as to insert multiple strings at the same time. This gives Algorithm
3, which is reduced to Algorithm 1 or 2 if we insert one string at a time.

When $B$ is represented by a balanced tree structure, the time complexity of
all three algorithms is $O(n\log n)$, where $n$ is the total number of symbols
in the input. However, we will see later that for short strings, Algorithm 3 is
substantially faster than the first two algorithms, due to the locality of
memory accesses, the possibility of cached B+-tree update, and the
parallelization of the `for' loop at line 1.  These techniques are more
effective for a larger batch of shorter strings.

Disregarding RLO/RCLO, Algorithm 3 is similar to BCR except that BCR
keeps $B$ in monolithic arrays. As a result, the time complexity of
BCR is $O(nl)$, where $l$ is the maximum length of reads, not scaling well to $l$.

\begin{algorithm}[ht]
\DontPrintSemicolon
\footnotesize
\KwIn{A string $P$ and an existing BWT $B$ for $T$}
\KwOut{BWT for $TP\$$}
\BlankLine
\textbf{Function} {\sc InsertIO1}$(B,P)$
\Begin {
	$c\gets\$$; $k\gets|\{i:B[i]=\$\}|$\;
	\For{$i\gets|P|-1$ \KwTo $-1$} {
		${\rm insert}(P[i],k;B_c)$\;
		$k\gets {\rm rank}(P[i],k;B_c)+|\{a<c,j:B_a[j]=P[i]\}|$\;
		$c\gets P[i]$
	}
	{\bf return} $B$
}
\caption{Append one string}
\end{algorithm}

\begin{algorithm}[ht]
\DontPrintSemicolon
\footnotesize
\KwIn{$B^{\rm RLO}(\mathcal{C})$ (or $B^{\rm RCLO}(\mathcal{C})$) and a string $P$}
\KwOut{$B^{\rm RLO}(\mathcal{C}\cup\{P\})$ (or $B^{\rm RCLO}(\mathcal{C}\cup\{P\})$)}
\BlankLine
\textbf{Function} {\sc InsertRLO1}$(B,P,{\it is\_comp})$
\Begin {
	$c\gets \$$\;
	\nl$[l,u)\gets \big[0,|\{i:B[i]=\$\}|\big)$\;
	\For{$i\gets|P|-1$ \KwTo $-1$} {
		$[l,u)\gets${\sc InsertAux}$(B,P[i],l,u,P[i+1],{\it is\_comp})$\;
	}
	{\bf return} $B$
}
\textbf{Function} {\sc InsertAux}$(B,c',l,u,c,{\it is\_comp})$
\Begin {
	$k\gets l$\;
	\If{is\_comp is {\bf true} {\bf and} $c'\not={\rm ``N"}$} {
		\For{$a=\$$ {\bf or} $c'<a<{\rm ``N"}$} {
			$k\gets k+\big[{\rm rank}(a,u;B_c)-{\rm rank}(a,l;B_c)\big]$\;
		}
	} \Else {
		\For{$\$\le a<c'$} {
			$k\gets k+\big[{\rm rank}(a,u;B_c)-{\rm rank}(a,l;B_c)\big]$\;
		}
	}
	${\rm insert}(c',k;B_c)$\;
	$m\gets|\{a<c,j:B_a[j]=c'\}|$\;
	{\bf return} $\big[{\rm rank}(c',l;B_c)+m,{\rm rank}(c',u;B_c)+m\big)$\;
}
\caption{Insert one string to RLO/RCLO-BWT}
\end{algorithm}

\begin{algorithm}[ht]
\DontPrintSemicolon
\footnotesize
\KwIn{Existing BWT $B$ and a list of strings $\{P_k\}_k$}
\KwOut{Updated BWT $B$ with strings inserted in the specified order}
\BlankLine
\textbf{Function} {\sc InsertMulti}$(B,\{P_k\}_k,{\it is\_sorted},{\it is\_comp})$
\Begin {
	\For{$0\le j<|\{P_k\}_k|$} {
		$A(j).c\gets \$$; $A(j).i\gets j$\;
		\If{is\_sorted is {\bf true}} {
			$[A(j).l,A(j).u)\gets [0,|\{i:B[i]=\$\}|)$\;
		} \Else {
			$A(j).l\gets A(j).u\gets|\{i:B[i]=\$\}|+j$\;
		}
	}
	$d\gets 0$\;
	\While{$|A|\not=0$} {
		\emph{Stable sort array $A$ by $A(\cdot).c$}\;
		\nl\For{$0\le j<|A|$} {
			$c\gets A(j).c$\;
			$A(j).c\gets P_{A(j).i}[|P_{A(j).i}|-1-d]$\;
			$[A(j).l,A(j).u)$\;
			\hspace{0.1cm}$\gets${\sc InsertAux}$(B,c,A(j).l,A(j).u,A(j).c,{\it is\_comp})$\;
		}
		\emph{Remove $A(j)$ if $A(j).c=\$$}\;
		$d\gets d+1$\;
	}
	{\bf return} $B$
}
\caption{Insert multiple strings}
\end{algorithm}
\end{methods}

\begin{table}[b]
\processtable{Performance of BWT construction}
{\footnotesize
\begin{tabular}{llcrrrl}
\toprule
Data$^1$& Algorithm&RCLO& Real  & CPU\% &RAM$^2$& Comments\\
\midrule
worm & nvbio       & -  & 316s  & 138\%&12.9G & See note$^3$\\
worm & ropebwt-bcr & -  & 480s  & 223\%&2.2G & -btORf\\
worm & Algorithm 3 & Yes& 506s  & 250\%&10.5G & -brRm10g \\
worm & Algorithm 3 & No & 647s  & 249\%&11.8G & -bRm10g \\
worm & beetl-bcr   & -  & 965s  & 259\%&1.8G & RAM disk$^4$\\
worm & beetl-bcr   & -  & 2092s & 122\%&1.8G & Network$^5$\\
worm & Algorithm 1 & -  & 5125s & 100\%&2.5G & -bRm0 \\
worm & beetl-bcrext& -  & 5900s &  48\%&0.1G & Network$^5$\\
12878&ropebwt-bcr  & -  & 3.3h & 210\%&39.3G & -btORf \\
12878&nvbio        & -  &4.1h  & 471\%&63.8G & See note$^6$\\
12878&Algorithm 3  & Yes&5.0h  & 261\%&34.0G & -brRm10g \\
12878&Algorithm 3  & No &5.1h  & 248\%&60.9G & -bRm10g \\
12878&beetl-bcr    & -  &11.2h & 131\%&31.6G & Network$^5$\\
Venter&Algorithm 3 & Yes& 1.4h & 274\%&22.2G & -brRm10g \\
Venter&Algorithm 3 & No & 1.5h & 274\%&22.8G & -bRm10g \\
mol  &Algorithm 3  & No & 6.8h & 285\%&20.0G & -bRm10g \\

\botrule
\end{tabular}}{$^1$Data sets --
{\it worm}: 66M$\times$100bp {\it C. eleganse} reads from SRR065390;
{\it 12878}: 1206M$\times$101bp human reads for sample NA12878~\citep{Depristo:2011vn}.
{\it Venter}: 32M$\times$875bp (in average) human reads by Sanger sequencing (\citealt*{Levy:2007uq}; http://bit.ly/levy2007);
{\it mol}: 23M$\times$4026bp (in average) human reads by Illumina's Moleculo sequencing (http://bit.ly/mol12878).
$^2$Hardware -- CPU: 48 cores of
Xeon E5-2697v2 at 2.70GHz; GPU: one Nvidia Tesla K40; RAM: 128GB; Storage:
Isilon IQ 72000x and X400 over network. CPU time, wall-clock time and peak
memory are measured by GNU time. $^3$Run with option `-R -cpu-mem 4096 -gpu-mem
4096'. NVBio uses more CPU and GPU RAM than the specified. $^4$Results and
temporary files created on in-RAM virtual disk `/dev/shm'. $^5$Results and
temporary files created on Isilon's network file system. $^6$Run with
option `-R -cpu-mem 48000 -gpu-mem 4096'.}
\end{table}

\vspace*{-1em}
\section{Results and Discussion}
We implemented the algorithm in ropeBWT2 and evaluated its performance
together with BEETL (http://bit.ly/beetlGH), the original on-disk
implementation of BCR and BCRext, ropeBWT-BCR (https://github.com/lh3/ropebwt),
an in-memory reimplementation of BCR by us, and NVBio (http://bit.ly/nvbioio), a
reimplementation of the CX1 GPU-based algorithm~\citep{DBLP:journals/corr/LiuLL14}.
Table~1 shows that for short reads (the worm and 12878 data sets), ropeBWT2 has comparable performance to
others. For the 875bp or so Venter data set, NVBio aborted due to insufficient
memory under various settings. We did not apply BCR because it is not designed
for long reads of unequal lengths. Only ropeBWT2 works with this data set and
the even longer moleculo reads.

In addition to fast construction, ropeBWT2 is able to add strings
to an existing BWT while maintaining RLO/RCLO. It is possible to delete
strings from a BWT and to generate a sampled suffix array by inserting
positions to a dynamic integer array in parallel, though these functionalities
have not been implemented yet.

\vspace*{-1em}
\section*{Acknowledgement}
\paragraph{Funding\textcolon} NHGRI U54HG003037; NIH GM100233

\bibliography{ropebwt2}
\end{document}